\documentclass[amsfonts, amssymb, amsmath, amsthm, pre, aps, nofootinbib, twocolumn,superscriptaddress]{revtex4-2}
\usepackage[english]{babel}
\usepackage[utf8]{inputenc}

\usepackage{amsthm}
\usepackage{mathtools}
\usepackage{physics}
\usepackage{xcolor}
\usepackage{graphicx}
\usepackage{adjustbox}
\usepackage{placeins}
\usepackage[T1]{fontenc}
\usepackage{csquotes}
\newtheorem*{theorem}{Theorem}

\usepackage[pdfstartview=FitH,
        breaklinks=true,
        bookmarksopen=false,
        bookmarksnumbered=true,
        colorlinks=true,
        linkcolor=black,
        citecolor=black,
        urlcolor=black,
        pdftitle={SymDel lookup for AIRR similarity search},
        pdfauthor={Touchchai Chotisorayuth, Yuta Nagano, Andreas Tiffeau-Mayer},
        ]{hyperref}

\begin{document}
\title{Lightning-fast adaptive immune receptor similarity search\\ by symmetric deletion lookup}

\author{Touchchai Chotisorayuth}
\affiliation{Department of Computer Science}
\author{Yuta Nagano}
\affiliation{Institute of Infection, Immunity and Transplantation}
\author{Andreas Tiffeau-Mayer}
%\email[Correspondence: ]{andreas.mayer@ucl.ac.uk}
\affiliation{Institute of Infection, Immunity and Transplantation}
\affiliation{Institute for the Physics of Living Systems \\ University College London}

\begin{abstract}
An individual's adaptive immune receptor (AIR) repertoire records immune history due to the receptors' antigen specificity.
Reading this record requires computational approaches for inferring receptor function from sequence, as the diversity of possible receptor-antigen pairs vastly outstrips experimental knowledge.
Identification of AIRs with similar sequence and thus putatively similar function is a common performance bottleneck in these approaches.
Here, we benchmark the runtime scaling of five algorithmic approaches to radius-based search for Levenshtein neighbors.
We show that a symmetric deletion lookup approach, originally proposed for spell-checking, is particularly scalable.
We introduce SymScan and XTNeighbor, optimized CPU and GPU software implementing a parallelized variant of the algorithm.
For one million input sequences, these tools identify all sequence pairs that differ by one or two edits in seconds, orders of magnitude faster than existing approaches.
We demonstrate how symmetric deletion lookup can be integrated as a pre-filtering step in T cell receptor metaclone discovery and B cell receptor lineage identification.
Our contribution is poised to greatly accelerate existing analysis pipelines and enable processing of immunosequencing data at scale.
\end{abstract}


\maketitle

Somatic recombination creates adaptive immune receptors (AIR) of immense diversity. Despite AIR generation being independent across individuals, convergent selection on shared antigens commonly produces similar antibodies \cite{Setliff2018MultiDonorLongitudinal,Raybould2021CurrentStrategies} and T cell receptors (TCRs) \cite{Venturi2008MolecularBasis,Elhanati2018PredictingSpectrum,Mayer2023MeasuresEpitope} -- so-called public responses. This fundamental insight has driven the emergence of computational approaches that leverage sequence similarity to generate insights from AIR repertoire data \cite{Shugay2015VDJtoolsUnifying,Madi2017CellReceptor,Dash2017QuantifiablePredictive,Mayer-Blackwell2021TCRMetaClonotypes,Pogorelyy2019FrameworkAnnotation,Valkiers2021ClusTCRPython,Chronister2021TCRMatchPredicting,compairr,ismart}.

While sequence-similarity searches have long been a mainstay of computational biology, approaches tailored to the particular data-distributional properties of AIR repertoires are beginning to emerge. Due to antigen-driven convergent selection and recombination biases repertoires often contain highly similar AIR sequences that differ by only a few edits in their hypervariable regions \cite{Dash2017QuantifiablePredictive,Madi2017CellReceptor,Mayer2023MeasuresEpitope}. The rationale for AIR similarity search also differs from traditional homologue search \cite{Pearson2013IntroductionSequence}, and instead exploits convergent selection on non-homologous proteins to predict function from sequence (for an example of similar ideas in the general protein bioinformatics literature, see \citet{Littmann2021EmbeddingsDeep}).

These tailored tools cover a number of different use cases: First, identification of sequence near-overlaps between samples can be used to define measures of repertoire similarity using software packages such as VDJtools \cite{Shugay2015VDJtoolsUnifying} and CompAIRR \cite{compairr}. Second, unsupervised clustering of AIR repertoires can identify receptor groups with putatively shared specificity -- commonly termed metaclones or specificity groups -- in both B cell receptors (BCRs) \cite{richardson_computational_2021} and T cell receptors (TCRs) using tools such as TCRdist \cite{Dash2017QuantifiablePredictive,Mayer-Blackwell2021TCRMetaClonotypes}, Gliph \cite{Glanville2017IdentifyingSpecificity,gliph2}, iSMART \cite{ismart} and ClusTCR \cite{Valkiers2021ClusTCRPython}.
Third, the identification of similar sequences can be used for the supervised prediction of AIR specificity by annotation transfer using tools such as VDJMatch \cite{Pogorelyy2019FrameworkAnnotation}, TCRMatch \cite{Chronister2021TCRMatchPredicting} and KA-Search \cite{Olsen2023KASearchMethod}.

\begin{figure*}
\centering
\includegraphics[width=0.8\textwidth]{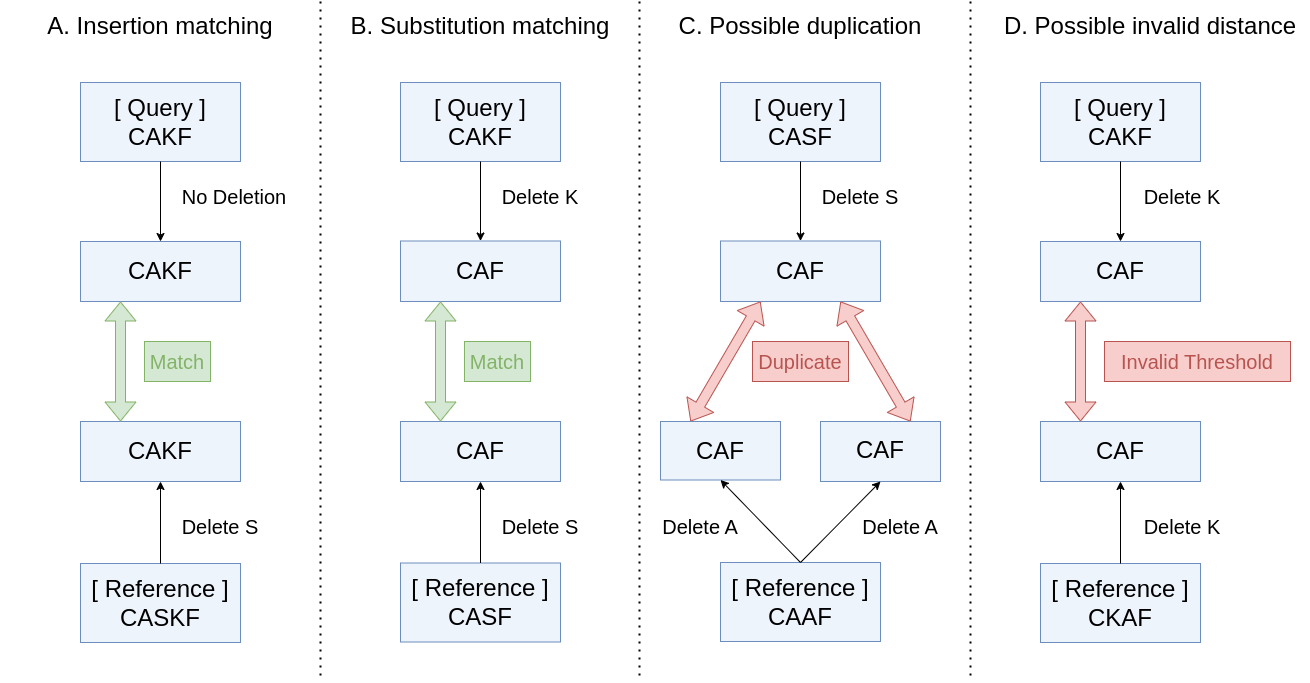}
\caption{
{\bf Symmetric deletion lookup by examples.} 
(A) Insertions (deletions) are identified by matches between a query (reference) and a deletion variant of a reference (query).
(B) Substitutions are identified by matches between deletion variants of both query and reference.
(C) Matches between multiple deletion variants of the sequence pair can lead to duplications.
(D) Matches between deletion variants can sometimes exceed the threshold distance ($d=1$ in this example).
Such false positives as shown in C and D can be filtered with post-processing verification.
}
\label{fig:symspell_case}
\end{figure*}

Across applications, fast identification of highly similar sequences is widely considered an important performance bottleneck. Combinatorial lookup algorithms have recently been proposed as a potential solution \cite{compairr,Valkiers2021ClusTCRPython}, which generate all possible variants of a query sequence below the threshold distance and then use exact hash-based comparisons to identify matches. This approach greatly reduces the computational complexity of finding single-edit neighbors compared to exhaustive search, but scales poorly to higher threshold distances.

In this paper, we demonstrate that a symmetric deletion (SymDel) lookup algorithm substantially extends the range of feasible sequence similarity thresholds. Our first contribution is to adapt a symmetric deletion approach \cite{symspell,bocek2007fast} from spell checking to AIR similarity search, and to benchmark it against other exact algorithms across a range of distance thresholds. Our second contribution is the implementation of SymScan and XTNeighbor, efficient parallelized variants of this algorithm. Both replace the multi-value hash map of the original algorithm with a sort-merge join, and XTNeighbor additionally streams the data and uses provably lossless prefix truncation for memory optimization. Finally, we demonstrate how this core algorithm supports a general identify-and-filter strategy applicable to many immunoinformatics use cases. Taken together, these advances provide a template for making AIR similarity search lightning fast.

The remainder of the paper is structured as follows: Section \ref{sec_symdel} introduces the SymDel lookup algorithm and our work on optimizing this algorithm. Section \ref{sec_results} compares the algorithmic complexity of different approaches and benchmarks the performance of our implementations on relevant use cases. In the final section, we discuss applications, limitations and future directions.

\section{Symmetric deletion lookup}
\label{sec_symdel}

\subsection{The symmetric deletion trick}

SymDel lookup is a variation of the combinatorial lookup algorithm that uses exact matching of only deletion variants on both the query and reference sides \cite{bocek2007fast,symspell}. This algorithm is surprisingly little known within the bioinformatics literature and we therefore review how it operates through a collection of examples provided in Fig.~\ref{fig:symspell_case}. From the SymDel perspective, generating insertions on the query side is equivalent to generating deletions on the reference end of the comparison (Fig.~\ref{fig:symspell_case}A). Similarly, generating substitutions is equivalent to performing deletion on both sides with the substituting character being removed (Fig.~\ref{fig:symspell_case}B).

By focusing only on deletions, the SymDel algorithm achieves significant efficiency gains compared to combinatorial lookup which requires the generation of all substitution-insertion-deletion variants.
For example, an amino-acid string of length 10 has 79,745 possible neighbors at Levenshtein distance 2 but only 55 deletion variants.
Importantly, deletion variant lookups do not miss true neighbor pairs: whenever two sequences lie within Levenshtein distance $d$, their sets of deletion variants up to depth $d$ intersect \cite{bocek2007fast}. 

An attentive reader might have noticed that SymDel can return duplicated (Fig.~\ref{fig:symspell_case}C) or irrelevant matches (Fig.~\ref{fig:symspell_case}D), as consecutive character repeats and character swaps may produce false positives. Fortunately, the number of returned candidate pairs is typically very small compared to all possible pairs that would be tested by a brute-force approach. Therefore, the output of the algorithm can be filtered in a post-processing step, which adds minimal overhead at the dataset sizes and distance thresholds typical of AIR applications. This cost does eventually come to dominate the total runtime as the number of candidate pairs grows superlinearly with input size, which we quantify in Appendix~\ref{app:scaling} (Table~\ref{tab:in-depth}).

\subsection{Overcoming parallelization and memory bottlenecks}
\label{parallel_bottleneck}

We next aimed to parallelize the SymDel algorithm to enable use of multi-core Central Processing Units (CPUs) or Graphics Processing Units (GPUs).
The original algorithm uses a multi-value hash map to collate input strings that share deletion variants.
Na\"{\i}ve implementations of this data structure are not well suited to parallelization, as multiple independent processes could attempt to modify the same entry.
To overcome this limitation, we replaced the multi-value hash map with a sort-merge join approach \cite{sort-merge-join}. This replacement trades a logarithmic factor in asymptotic complexity against increased simplicity compared to concurrent hash maps.

At a high level, the algorithm proceeds as follows:
First, generate an array containing all tuples of deletion variants and original strings for both the query and reference sets.
Second, sort those tuples by the deletion variant.
This guarantees that all strings with matching deletion variants will be stored contiguously.
Third, generate all candidate matches and subsequently deduplicate and filter them as in the standard SymDel approach.

To showcase the computational capability of this modified SymDel algorithm we implemented SymScan and XTNeighbor. SymScan is a general-purpose, CPU-based tool for string similarity search in Rust complemented with Python bindings. XTNeighbor is a GPU-based implementation, used through a command-line interface rather than a library API. All results reported here use its streaming variant, which we refer to simply as XTNeighbor. Step-by-step procedures for both tools are provided in Appendix~\ref{app_sup_methods}, which also describes the XTNeighbor base algorithm that the streaming variant builds on. Both tools support arbitrary edit distance thresholds. SymScan places no limit on sequence length, while XTNeighbor accepts sequences of up to 500 characters, well beyond the length of full adaptive immune receptor variable regions. 

As datasets are scaled to millions of sequences, memory also becomes a bottleneck (see  Fig.~\ref{fig:symscan_memory_scaling}): the number of deletion variants grows linearly with dataset size, with a prefactor depending on average sequence length and the distance threshold (see Appendix~\ref{app:scaling}). One way to reduce memory consumption is to minimize the memory taken up by data structures in the algorithm. For example, deletion variants can be stored as (smaller) hashes rather than full strings during the sort-merge join. This is an optimization employed in SymScan. XTNeighbor achieves a similar effect by left-aligning and truncating input sequences to 18-character prefixes for the purpose of candidate generation, while retaining the full sequences for the final verification step. Prefix truncation does not yield any false negatives, as shown in Appendix~\ref{app:truncation}. Both methods trade an increased rate of false positive candidate pairs for lower memory use, which are removed in a final filtering step by comparison of the full sequences. We will show additional applications of this identify-and-filter pattern in Section~\ref{sec:identify_and_filter}. To further reduce memory consumption, XTNeighbor streams the data, using algorithmic extensions described in Appendix~\ref{app:streaming}.

\section{Results}
\label{sec_results}

\begin{figure*}
\includegraphics[width=\textwidth]{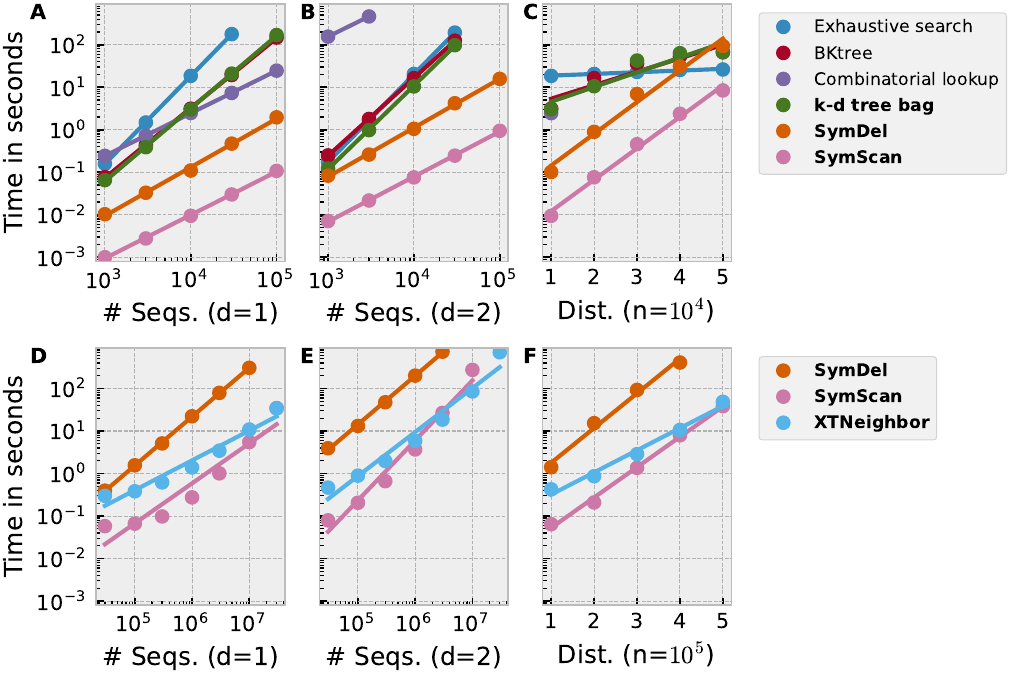}
\caption{{\bf Benchmark of similarity search algorithms.} Time measured in seconds for near-sequence neighbor search as a function of input size for various algorithms and Levenshtein distance threshold. (A-C) Comparison of CPU-based algorithms on medium-sized data. (D-F) Comparison of GPU-based algorithm and most scalable CPU-based approaches on large-sized data. 
 (A, D) $d=1$ and (B, E) $d=2$. (C, F) Side-by-side comparison of search time at different thresholds for 10,000 and 100,000 sequences, respectively. In bold: Algorithms that have not previously been applied to AIR similarity search to our knowledge. Lines are linear fits on log-log scale to assess scaling deviations. Parameter settings exceeding 1000s of run time are excluded from the plot.}
\label{fig:alg_benchmark}
\end{figure*}

\subsection{Computational complexity of algorithmic approaches to AIR similarity search}

Existing AIR tools rely on a range of algorithmic approaches, and it remains unknown which approach for exact radius search scales best to cohort-scale repertoire sizes.
We thus set out to comprehensively benchmark algorithmic approaches to AIR similarity search on an equal footing. Our benchmark addresses how algorithm speed scales with the size of the input data and the similarity threshold choice. By studying empirical scaling in the relevant ranges of dataset sizes we can compare algorithms without being overly biased by implementation details, as the scaling exponent is largely insensitive to constant factors.

We identified three alternative approaches from a literature search for exact Levenshtein distance radius-based similarity search on strings:
\begin{itemize}
\item Exhaustive search: the brute-force algorithm comparing all pairs of sequences
\item Combinatorial lookup: a hashing-based approach that generates all possible neighboring sequences \cite{compairr}
\item BKtree: a tree-based edit distance neighbor search algorithm for strings \cite{bktree}
\end{itemize}
For the BKtree algorithm we made use of the bktree library \cite{bktree_lib}. For all other algorithms we used our own Python implementations.

Sequence-to-vector representation approaches have gained tremendous popularity in bioinformatics applications over the last decade \cite{rep1,dong2019learning,rep3,rep4}. We thus  additionally implemented a representation-based algorithm (described in detail in Appendix \ref{seqtovec}) that allows exact Levenshtein distance radius searches as an exemplar of this paradigm:
\begin{itemize}
\item k-d tree bag: a k-d tree-based neighbor search algorithm for vectors, which we applied to identify similar AIRs using a novel bag-of-amino-acids representation that lower bounds edit distance (introduced in Appendix \ref{seqtovec}).
\end{itemize}

We benchmarked the task of finding all neighbor pairs within a given set of sequences. 
To make our benchmark realistic, we used real-world sequencing data of TCR$\beta$ complementarity determining region 3 (CDR3) sequences from a large cohort of healthy human subjects \cite{Emerson2017}. We created random subsets of the data of the following sizes: 1k, 3k, 10k, 30k, and 100k sequences. 
Fig.~\ref{fig:alg_benchmark}A shows the runtimes for different algorithms at a Levenshtein distance threshold of $d=1$ (single-edit neighbors). Fig.~\ref{fig:alg_benchmark}B shows the corresponding results for $d=2$ (two-edit neighbors). Fig.~\ref{fig:alg_benchmark}C compares the runtime of all algorithms across $d=1,2,3,4,5$ for 10k sequences.

\begin{table}[b]
\centering
\begin{tabular}{|p{3.5cm}|p{1.5cm}|p{1.5cm}|} 
\hline
Algorithm & $d=1$ & $d=2$  \\
\hline
Combinatorial Lookup & 1.00 & 0.99 \\
SymDel Lookup & 1.14 & 1.15 \\
SymScan & 1.02 & 1.06 \\
BKtree & 1.64 & 1.83 \\
k-d tree bag & 1.71 & 1.96 \\
Exhaustive search & 2.07 & 2.06 \\
\hline
\end{tabular}
\caption{Empirical time complexity of algorithms on the mid-size dataset benchmark (Fig.~\ref{fig:alg_benchmark}A,B) as determined by power-law fits for Levenshtein distance thresholds of $d=1$ and $d=2$.}
\label{tab:scalability}
\end{table}

To assess empirical scalability, we fitted each curve to a power-law,  $\log(t) = A \log(n) + B$, where  $t$ denotes run time, $n$ denotes number of inputs, and $A$/$B$ are fitted parameters. The resulting exponents $A$, reported in Table \ref{tab:scalability}, are intended to provide a comparative overview of algorithmic efficiency in a practically relevant regime. Note that as input sizes go to infinity, the scaling of all algorithms is expected to crossover into a quadratic regime, as the number of neighbors grows with $n^2$.

As expected, the exhaustive search time scales $\sim O(n^2)$, as this algorithm considers all $n(n-1)/2$ pairs of sequences. Combinatorial lookup reduces empirical time complexity to $\sim O(n)$, matching theoretical expectations: it performs $O(n)$ hash-table queries, each taking $O(1)$ time. SymDel similarly has close-to-linear scaling, demonstrating that for tested sample sizes, most time is spent on the neighbor candidate generation step (with expected $O(n)$ scaling) rather than on the post-processing (with expected $O(n^2)$ scaling). The sort-merge-join approach implemented in SymScan exhibits the same empirical scaling, but has a smaller prefactor due to the more efficient implementation. BKtree and k-d tree show similar performance, with scaling exponents intermediate between the other approaches. 

While both combinatorial and SymDel lookups have a time complexity $\sim O(n)$, performance depends very differently on the threshold choice $d$ (Fig.~\ref{fig:alg_benchmark}C). SymDel is only $\sim$10x slower at $d=2$, while the combinatorial approach already exceeds the maximal computational time. This more favorable scaling translates into an expansion of the range of sequence similarity thresholds for which lookup-based approaches remain feasible. Indeed, in our comparison SymScan remains the fastest method up to $d=5$, although its margin over exhaustive search shrinks from $\sim$2000-fold at $d=1$ to $\sim$3-fold at $d=5$.

The SymDel approach still has its bottlenecks.
One bottleneck lies in the number of generated deletion variants: for sequences of length $L$ and edit threshold $d$, SymDel generates $\sum_{i=1}^{d}\binom{L}{i}$ deletion variants per sequence.
Although the number of variants grows combinatorially with $d$, AIR CDR3 sequences are typically short ($L \approx 10$--$20$ amino acids) and biologically relevant edit thresholds are commonly modest (e.g., $d \leq 2$), keeping candidate generation tractable in practice.
A second bottleneck arises from the increase in the expected number of candidate pairs.
If a deletion variant is shared by $m$ sequences, this will generate $m(m-1)/2$ candidate pairs.
Typical values of $m$ will grow with $n$ and $d$, and the time spent filtering the candidate pairs can  become substantial.
The empirical consequences of both bottlenecks are analyzed in Appendix~\ref{app:scaling}. 
To reduce this, users might want to collapse exact duplicates before analysis where they exist, as they would give rise to many shared deletion variants.
Section~\ref{sec:identify_and_filter} revisits these issues and introduces a restriction of SymDel candidate generation to short k-mers  to replace searching at large thresholds on the full string.

\subsection{Benchmarking performance of optimized implementations}

We next used the same dataset to benchmark the optimized and parallelized implementations of symmetric deletion in SymScan (CPU-only) and XTNeighbor (GPU-accelerated) at larger scale. To this end we expanded experiments to larger samples, considering 30k, 100k, 300k, 1M, 3M, 10M, and 30M sequences. In contrast to the previous benchmark, we now used 8 CPU cores for SymScan to test its performance gain from parallelization. Fig.~\ref{fig:alg_benchmark}D-F shows the results of this benchmark.

SymScan and XTNeighbor reduce computing time by 10x-100x compared to the vanilla Python SymDel implementation. To determine the origin of this speedup, we ran SymScan with a varying number of CPU cores. SymScan achieves a $\sim$6x speedup when going from a single to 8 cores (Fig.~\ref{fig:symscan_ncpu_scaling}), which shows how the algorithm exploits multi-core processing, while the remaining speedup with respect to SymDel stems from the more optimized implementation. SymScan identifies all possible two-edit neighbors ($d=2$) in only $\sim$4s among 1,000,000 sequences, which puts interactive analysis of bulk repertoire sequencing datasets within reach.

Larger similarity thresholds remain feasible but become increasingly costly: at 100,000 sequences, the run times grow by roughly 3--6x for each additional edit (Fig.~\ref{fig:alg_benchmark}F), reflecting the increasing number of deletion variants and candidate pairs (Appendix~\ref{app:scaling}), and the Python SymDel implementation exceeds our 1000\,s limit at $d=5$. In addition, SymScan is substantially more memory efficient than the Python SymDel implementation, as shown in Fig.~\ref{fig:symscan_memory_scaling}. XTNeighbor removes memory constraints even more directly by streaming the data. Although it is slower than SymScan at small input sizes owing to GPU overheads, it is the tool that is most scalable to the very largest datasets.

Interestingly, empirical scaling of SymDel and SymScan remains sub-quadratic across the fitted range (Table~\ref{tab:scalability_gpu}), with scaling exponents only approaching 2 at the largest dataset sizes. This puts the onset of quadratic scaling at the upper end of the tested range.

\begin{table}[h]
\centering
\begin{tabular}{|p{3.5cm}|p{1.5cm}|p{1.5cm}|} 
\hline
Algorithm & $d=1$ & $d=2$  \\
\hline
SymDel Lookup & 1.15 & 1.14 \\
SymScan & 0.94 & 1.41 \\
XTNeighbor & 0.70 & 1.03 \\
\hline
\end{tabular}
\caption{Empirical time complexity of algorithms on the large dataset benchmark (Fig.~\ref{fig:alg_benchmark}D,E) as determined by power-law fits for Levenshtein distance thresholds of $d=1$ and $d=2$.}
\label{tab:scalability_gpu}
\end{table}

\subsection{Application examples} \label{sec:identify_and_filter}

In a first application, we computed repertoire overlaps across a cohort of $M$ repertoires. We again used data from Emerson et al. \cite{Emerson2017}, and calculated all pairwise overlaps for $M=2, 4, 8, 16, 32, 64$. To support this application, we extended SymScan and XTNeighbor with a wrapper that aggregates an $M \times M$ overlap matrix from identified neighbor pairs, where each entry contains the number of neighbors between two repertoires. This task allowed us to compare our approach to a recently published tool, CompAIRR \cite{compairr}, which also performs such computations, but uses combinatorial lookup based on a custom Bloom filter \cite{bloom}. At $d=1$, performance is similar across tools for both Hamming and Levenshtein distance metrics at small $M$, but the better scaling behavior of SymScan and XTNeighbor becomes apparent as $M$ increases (Fig.~\ref{fig:compairr}A,B). At $d=2$, SymScan and XTNeighbor are substantially faster than CompAIRR regardless of M (Fig.~\ref{fig:compairr}C), consistent with the increasing inefficiency of CompAIRR's combinatorial lookup at larger thresholds. We could not perform the corresponding comparison for Levenshtein distance at $d=2$ (Fig.~\ref{fig:compairr}D), as CompAIRR does not support this setting.

\begin{figure}
\includegraphics{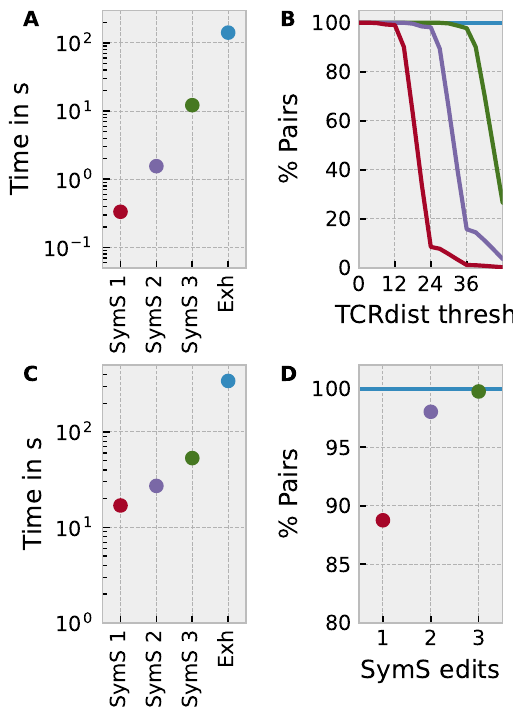}
\caption{{\bf SymScan as a candidate filter.} (A) Time measured in seconds for similarity search among 40,000 sequences using exhaustive TCRdist calculations (Exh) or a combination of SymScan and TCRdist filtering (SymScan with $d = 1,2,3$). (B) Percentage of TCR pairs within a given TCRdist threshold that are recovered by the SymScan pre-filtered pipeline, expressed relative to the pairs found by exhaustive TCRdist search. (C) Time measured in seconds for clonal lineage identification among BCR sequences in all VJ-length groups with more than 100 unique sequences in a donor from \cite{briney2019commonality}, using exhaustive Hamming distance calculation (Exh) or a combination of k-mer-based SymScan candidate identification and Hamming filtering (SymScan with $d = 1,2,3$). (D) Percentage of all BCR pairs with >85\% sequence similarity across the CDR3 nucleotide sequence that are recovered by SymScan pre-filtering, expressed relative to exhaustive Hamming distance search, as a function of the k-mer edit distance threshold.
In (B) and (D), colors denote the SymScan pre-filtering threshold -- $d=1$ (dark red), $d=2$ (purple), $d=3$ (green) -- matching the corresponding categories in (A) and (C). The horizontal blue line marks the exhaustive reference (Exh), which recovers 100\% of pairs by definition.
}
\label{fig:tcrdist_benchmark}
\end{figure}

We next asked how the algorithmic improvements in identifying similar sequences by Levenshtein distance might be leveraged in T cell metaclone discovery and antibody lineage clustering. We reasoned that in these applications SymScan could be used as a pre-filtering step to reduce the search space for neighbor lookups using a slower but more informative sequence metric.

For metaclone discovery, we combined SymScan with TCRdist \cite{Dash2017QuantifiablePredictive, Mayer-Blackwell2021TCRMetaClonotypes}, a TCR sequence metric that computes the distance between two receptors as a weighted edit distance using a modified BLOSUM62 substitution matrix, taking into account all CDRs 1, 2 and 3. Our combined SymScan-based pipeline uses only the CDR3 for pre-filtering, but considers the full TCR$\beta$ during filtering of candidates to retain accuracy. We benchmarked the combined algorithm against the out-of-memory sparse neighbor-finding implementation provided by the TCRdist3 Python package \cite{Mayer-Blackwell2021TCRMetaClonotypes}, using a dataset of 40,000 TCR$\beta$ sequences from a single donor, drawn from \cite{Emerson2017}.

Pre-filtering decreased runtime $\sim$100x and $\sim$10x at Levenshtein distance thresholds of $d = 2$ and $d = 3$, respectively (Fig.~\ref{fig:tcrdist_benchmark}A). To determine how many sequence pairs are missed with pre-filtering, we calculated the fraction of all pairs recovered by the combined pipeline. Recovery exceeded 95\% up to TCRdist thresholds of 12, 24, and 36 at $d=1$, $d=2$, and $d=3$, respectively (Fig.~\ref{fig:tcrdist_benchmark}B). These results suggest the combined approach can deliver practically relevant speed gains while preserving high accuracy for similarity thresholds commonly used for clustering \cite{lindeboom2024human,turner2026evolution}.

To demonstrate a use-case in clonal lineage identification in BCR sequencing data, we used a deeply sequenced heavy-chain BCR repertoire from \citet{briney2019commonality} comprising close to one million unique BCRs. Because somatic hypermutation preserves VJ gene usage and CDR3 length (in the absence of indels), we first stratified sequences into VJ-length groups, following previous work \cite{gupta2015change,spisak2024combining}. Within each group, we then identified pairs of sequences with >85\% sequence similarity across the CDR3 at the nucleotide level. Given an average CDR3 length of 50 nt, this threshold corresponds to 7 substitutions, which is beyond the direct reach of the symmetric deletion approach. We therefore applied SymScan to a short k-mer ($k=6$) centered in the CDR3, hypothesizing that k-mer-based candidate generation at $d=1,2,3$, followed by full-length Hamming distance filtering, would increase speed while recovering a substantial fraction of pairs.

We benchmarked this pipeline against exhaustive pairwise search across all VJ-length groups with more than 100 unique sequences. SymScan pre-filtering recovered >99.5\% of pairs at $d=3$ while running $\sim$6x faster than exhaustive search; smaller $d$ gave further speed gains at the cost of reduced recovery (98\% of pairs at a $\sim$13x speed-up for $d=2$). These results illustrate how k-mer-based pre-filtering might extend the symmetric deletion trick to settings with longer sequences and larger similarity thresholds, where a na\"{\i}ve application would not be feasible.

\section{Discussion}

In this work, we benchmarked similarity search algorithms for use in AIR repertoire analysis, and identified symmetric deletion as a particularly well-suited approach.
We further proposed a parallelized variant of SymDel lookup, which replaces hash-based lookups with a sort-merge-join approach.
Our implementations of this modified algorithm (SymScan and XTNeighbor) scale to cohort-size AIR repertoire data and provide meaningful speed-ups over alternative approaches.
In practice, we recommend SymScan as the default: it runs on CPUs, can be used from Python, and was faster in our benchmarks on datasets of up to about a million sequences. XTNeighbor is preferable for the largest workloads when a GPU is available and where its streaming design keeps memory use constant.

Our algorithmic advance can be incorporated into existing analysis pipelines for increased speed and scalability.
An important application is AIR clustering to discover groups of putatively antigen-specific AIRs in an unsupervised manner \cite{Glanville2017IdentifyingSpecificity,Dash2017QuantifiablePredictive,gliph2,Valkiers2021ClusTCRPython,ismart,Mayer-Blackwell2021TCRMetaClonotypes}. In recently published work, we have already built on the symmetric deletion approach to develop Metaclonotypist, as a highly scalable clustering tool, which allowed us to identify generalisable \textit{Mycobacterium tuberculosis}-reactive and HLA-restricted T cell metaclones from \textit{in vivo} tuberculin skin test responses across a cohort of donors \cite{turner2026evolution}. Fast candidate generation prior to TCRdist computation was essential to make accurate clustering tractable at the size of the established cohort, illustrating how the presented algorithmic advances can enable immunological discovery. Another application for the SymDel approach is fuzzy database lookup \cite{Shugay2015VDJtoolsUnifying,Chronister2021TCRMatchPredicting}, where entire patient AIR repertoires can be queried against databases of receptors with known target specificities or disease associations. We note that while we have used TCR$\beta$ chain data for the majority of benchmarks, symmetric deletion lookup is also applicable to antibody-paratope clustering \cite{richardson_computational_2021} and paratope-based antibody database matching \cite{Olsen2023KASearchMethod}.

The approach has limitations at large thresholds and for long sequences. Candidate pair generation, deduplication, and distance verification become dominant costs at larger thresholds, motivating strategies that restrict comparisons to short sequence fragments. Prefix restriction is provably exact (Appendix~\ref{app:truncation}), whereas k-mer-based restriction trades exactness for speed. We demonstrated the latter in Section~\ref{sec:identify_and_filter} for antibody clonotyping, where identifying clonal families typically requires full-length nucleotide comparison \cite{spisak2024combining}; restricting to a short k-mer within the CDR3 instead recovered most pairs at substantially higher speed. Somatic hypermutation hotspots could bias which pairs a k-mer filter recovers, a risk left to future work.

Our benchmarking has been limited to existing approaches that support exact distance-based similarity search, excluding potential speed gains attainable by heuristic neighbor search (e.g., ClusTCR \cite{Valkiers2021ClusTCRPython}, built on the Faiss library \cite{faiss}) or pre-alignment (e.g., KA-Search \cite{Olsen2023KASearchMethod}).
By making exact Levenshtein distance search at scale feasible, our work will enable future investigation of the accuracy-speed trade-offs inherent in relaxations of exact search. Such comparisons might also extend to ongoing efforts to accelerate similarity search through approximate, embedding-based approaches, using windowed k-mer embeddings \cite{Yang2026tcrseek}, vectorized approximations of TCRdist \cite{Valkiers2026vectorized} or protein language model embeddings \cite{nagano2025contrastive}.

With the steady advancement in throughput of immunosequencing pipelines -- including large-scale repertoire profiling at the single-cell level \cite{sureshchandra2025deep,xue2025pan} -- we anticipate that scalable computational methods such as SymScan will become increasingly important to extract immunological insights from repertoire data. Fast mining of sequence neighbor pairs from repertoires using SymScan could also provide new training data for weakly supervised machine learning approaches \cite{ruffolo2021decipheringantibody}, which might help overcome limitations with currently available labeled data \cite{hudson,hummer2025investigating}. Finally, given the broad uses of sequence similarity search across bioinformatics the SymDel approach may prove useful to domains beyond immunology \cite{steinegger2017mmseqs2,mahe2022swarm}. To support such uses, SymScan is available as a general-purpose tool for string similarity search, which can be applied to sequence data from any domain.

\FloatBarrier

\section{Methods}

\subsection{Benchmarking details}
\label{sec_bench_details}

Benchmarks were performed on a Linux machine with an Intel i9-13900 and 128 GB of RAM. GPU benchmarks used a NVIDIA GeForce RTX 4090 with 24 GB of VRAM running CUDA 13.0. 
To obtain reproducible timing measurements we controlled process placement and processor frequency. Each benchmark process was bound with taskset to one logical CPU per physical performance core, so that no two threads shared a core's execution units or L1/L2 cache, and efficiency cores (which differ in performance characteristics) were never used. This restricted CPU-parallelism to 8 cores on this system. Turbo Boost was disabled as it is load- and duration-dependent, and might thus add uncontrolled variability to timing measurements.

At each dataset size, run times were averaged over 10 randomly chosen samples. Parameter combinations where a single runtime exceeded 1000s were excluded from the benchmark to maintain a reasonable total runtime.

All algorithmic benchmarks reported in Fig.~\ref{fig:alg_benchmark}A-C were performed on a single CPU core. The larger-scale benchmarks in Fig.~\ref{fig:alg_benchmark}D-F were performed on 8 CPU cores for SymScan, as were all application examples in section~\ref{sec:identify_and_filter}. CompAIRR was also run on 8 CPU cores using \texttt{-t 8}, and compared against the symscan-airr command line interface. XTNeighbor was run on a single GPU.

\subsection{Implementation validation}
\label{sec_validation}

We verified that all implementations return identical results. On sets of 5,000 sequences at thresholds $d=1,2,3$, the neighbors returned by the SymDel implementation in pyrepseq, SymScan, and XTNeighbor were compared for exact equality against a brute-force double loop over all sequence pairs. At scales where brute-force comparison is no longer feasible, SymScan and XTNeighbor were cross-checked against each other on up to $3\times10^6$ sequences at $d=1$, $10^6$ sequences at $d=2$, and $10^5$ sequences at $d=3$. All comparisons agreed exactly, consistent with the exactness guarantee for prefix-truncated candidate generation used in XTNeighbor (Appendix~\ref{app:truncation}). As an additional robustness check, we also confirmed equivalence under random shuffling of inputs, extending this check for XTNeighbor to $10^7$ sequences. Repertoire overlap matrices computed with SymScan and XTNeighbor were furthermore identical to those computed by CompAIRR for all settings supported by CompAIRR. The notebooks performing these checks are provided in the benchmarking repository.

\subsection{Data curation}

The TCR sequences used for benchmarking were drawn from \citet{Emerson2017}. The dataset was filtered to retain only functional and deduplicated sequences. The dataset was filtered to remove CDR3 sequences with a length <5 or >30 amino acids (retaining $>$99\% of sequences).

The BCR sequences used for benchmarking were drawn from \citet{briney2019commonality}.
The dataset was filtered to retain only sequences with a length >6 and <90 nucleotides (retaining $>$99\% of sequences).

\vspace{5pt}

{\bf Code Availability.} SymScan is available at \url{https://github.com/yutanagano/symscan}. It includes both a general purpose interface and an command-line tool (symscan-airr) for adaptive immune receptor repertoire comparisons. Code to reproduce the benchmarks and the XTNeighbor implementations are available at \url{https://github.com/heartnetkung/XT-neighbor}. A Python implementation of SymDel for AIR sequence analysis applications is available as a part of Pyrepseq at \url{https://github.com/andim/pyrepseq}. 

\vspace{5pt}

{\bf Data Availability.} Raw data from \citet{Emerson2017} are available at  \url{https://clients.adaptivebiotech.com/pub/emerson-2017-nat-genet}. Raw data from \citet{briney2019commonality} are available at \url{https://www.ncbi.nlm.nih.gov/bioproject/406949}. The benchmarking repository provides scripts to download the raw TCR data and regenerate the filtered benchmarking datasets, and includes the filtered BCR data.

\vspace{5pt}

{\bf Acknowledgements.} The authors thank Martina Milighetti, James Henderson, and Rishika Saxena for discussion and beta-testing. The work was supported in part by funding from the Wellcome Leap HOPE Program.

\bibliography{references}

\clearpage
\onecolumngrid
\appendix

\renewcommand{\thefigure}{S\arabic{figure}}
\setcounter{figure}{0} 
\renewcommand{\thetable}{S\arabic{table}}
\setcounter{table}{0}

\section{Supplementary figures}

\begin{figure}[h!]
\centering
\includegraphics{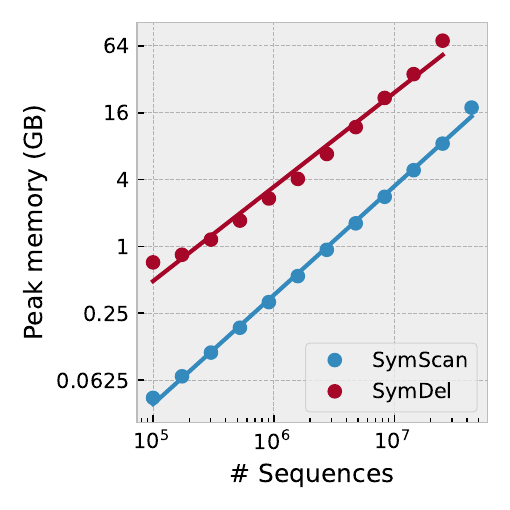}
\caption{
{\bf Memory scaling of SymScan and SymDel.}
Peak memory usage of SymScan and SymDel as a function of the number of input sequences for Levenshtein distance threshold $d=1$. Memory usage of SymScan scales linearly with input size (power-law exponent $\approx$1.0), while SymDel scales weakly sublinearly ($\approx$0.85, due to repeated deletion variants) but with a substantially larger prefactor. SymScan's lower memory footprint is aided by storing deletion variants as small hashes rather than full strings (Section~\ref{parallel_bottleneck}). At this threshold, SymScan can process tens of millions of sequences within the 32 GB of RAM of a typical workstation.
}
\label{fig:symscan_memory_scaling}
\end{figure}

\begin{figure}[h!]
\centering
\includegraphics{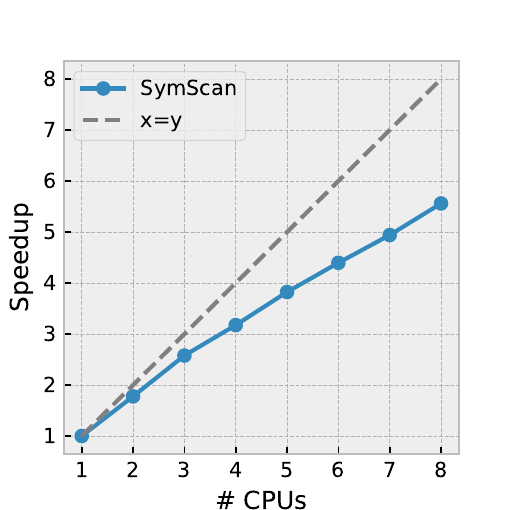}
\caption{
{\bf CPU parallelization scaling of SymScan.}
Multi-core speedup of SymScan relative to single-core runtime for 1M input sequences at Levenshtein distance threshold $d=2$. The dashed line indicates linear speedup (upper bound).
}
\label{fig:symscan_ncpu_scaling}
\end{figure}

\begin{figure}[h]
    \includegraphics{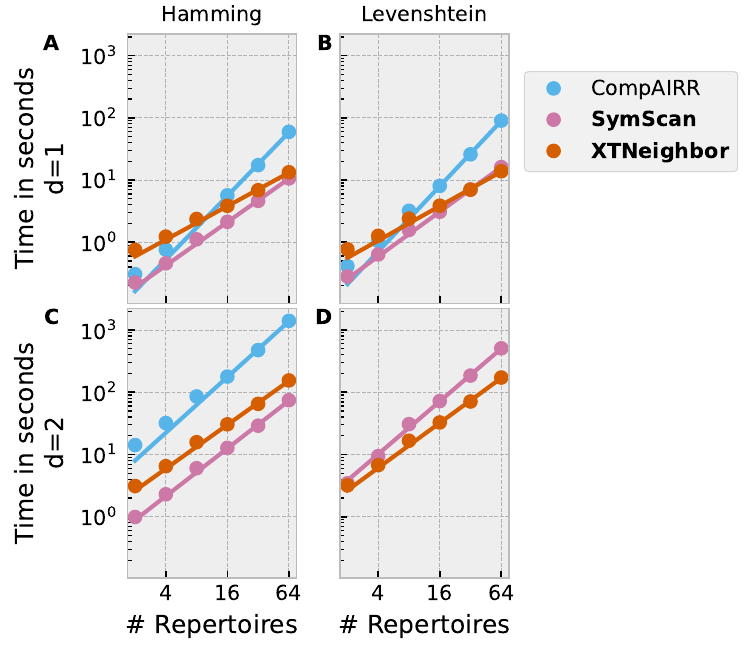}
    \caption{{\bf Benchmark of repertoire overlap computation speed against CompAIRR.} Time measured in seconds as a function of the number of input repertoires for (A) Hamming distance, $d=1$, (B) Levenshtein distance, $d=1$, (C) Hamming distance, $d=2$, and (D) Levenshtein distance, $d=2$. CompAIRR is not shown in (D) as it does not support Levenshtein distance search for $d>1$.}
    \label{fig:compairr}
\end{figure}

\clearpage

\section{Prefix truncation recovers all similar pairs within the edit distance threshold}
\label{app:truncation}

Let $u, v$ be two strings, and let $u'$ and $v'$ denote the prefix-truncated versions of $u$ and $v$ obtained by removing any characters exceeding a length of $L$. We will denote by $M = \max(|u|, |v|)$ the length of the longer string, and focus on the non-trivial case where $M \geq L$. We show that if $u$ and $v$ are within Levenshtein distance $\leq d$, then $u'$ and $v'$ share a common deletion variant of length at least $L-d$.

To show this, we consider an optimal alignment of $u$ and $v$. In such an alignment, pairs of aligned matched characters form a common subsequence $c$ of length at least
\begin{equation}
|c| \geq M - d.
\end{equation}
We next restrict $c$ to pairs within the first $L$ characters of $u$ and $v$. Because the matched positions are increasing, any discarded pairs form a suffix of $c$, implying that at most
\begin{equation}
M-L
\end{equation}
pairs will be discarded. Therefore, the length of the resulting subsequence $c'$ is at least
\begin{equation}
|c'| \geq |c| - (M-L) \geq M - d - (M-L) = L - d.
\end{equation}
Because $c'$ is a common subsequence of $u'$ and $v'$, it follows that $u'$ and $v'$ share a common deletion variant of length at least $L - d$. Since $|u'|, |v'| \leq L$, $c'$ can be reached from each sequence by at most $d$ deletions, and is therefore among the deletion variants generated at threshold $d$.

In practice, this means that restricting candidate generation to a prefix does not compromise the exactness of the algorithm, as long as the prefix length exceeds the edit distance threshold. We note that in contrast to the lemma shown above the Levenshtein distance between $u'$ and $v'$ may exceed $d$. Consider the strings CAT and COAT. Their edit distance is 1, but the edit distance between their prefix-truncated versions CAT and COA is 2. However, the prefix-truncated strings still share a common deletion variant of length 2 (CA), which is sufficient for candidate generation.

\section{Extension of the algorithm to enable streaming}
\label{app:streaming}

At very large input sizes memory requirements can become limiting. A streaming implementation of the algorithm provides a means to reduce working memory needs. In a streaming approach, the input is divided into chunks which are processed individually. XTNeighbor implements such a streaming approach on the GPU, allowing it to operate on inputs of arbitrary size with constant working memory needs.

To allow streaming, two problems with the original parallelized algorithm need to be overcome: directional change and unbounded matching. Directional change occurs during the grouping operation, where the direction of the processed data is changed from its original order to a new grouped order. Unbounded matching occurs during duplication removal, which requires the matching of duplicated values across chunks. Neither operation can be performed directly in a streaming fashion, as values from the current chunk would need to be retained for possible matching in future chunks. To solve both problems, we make use of a 2-dimensional buffer which stores data in a sorted grid format such that data written in a row-wise manner can be accessed in a column-wise manner in another order (Fig. \ref{fig:data_flow}). Such a buffer solves the directional change problem, if one of the access directions is in the grouping order. Furthermore, by binning the data into columns consistently, we can ensure that, regardless of chunk, a particular sequence is always assigned to the same column. If all repeated values are part of the same column, the unbounded matching problem is reduced to column-bounded matching, the latter of which is feasible with constant memory size.

\section{Empirical scaling behavior of the symmetric deletion approach}
\label{app:scaling}

To gain insight into the memory requirements and speed bottlenecks of the symmetric deletion approach, we calculated the number of entries considered at different steps in the algorithm on a typical dataset. We also benchmarked the run time of each step (variant generation, match candidate generation, deduplication, distance-based filtering) using XTNeighbor, which separates these four steps into distinct processing streams. As with the other benchmarks reported in the main text, we used TCR$\beta$ CDR3 sequences drawn from the Emerson set \cite{Emerson2017} for these experiments. Note that this benchmark was performed on a Google Colab instance with a T4 GPU, using a version of the dataset that excludes CDR3 sequences longer than 18 amino acids, and timings are reported as averages over 10 runs.

\begin{table}[h!]
\begin{tabular}{lllllllllll}
\hline 
Distance & \multicolumn{6}{c}{Number of entries} & \multicolumn{4}{c}{Run times in seconds} \\
& \texttt{Inputs} & \texttt{AllPairs} & \texttt{DeletionVariants} & \texttt{MatchedPairs} & \texttt{UMatchedPairs} & \texttt{Outputs} & S1 & S2 & S3 & S4\\
\hline
1 & 100K & 5B & \underline{1.6M} & 174.8K & 96.0K (55\%) & 3.7K (4\%)     & <0.1 & <0.1 & <0.1 & <0.1\\
 & 1M &  500B & \underline{15.8M} & 3.0M & 2.0M (67\%) & 370.4K (19\%)       & 0.3 & \underline{0.4} & 0.2 & <0.1\\
 & 10M & 50000B & \underline{158.4M} & 156.5M & 122.0M (78\%) & 36.9M (30\%)  & 2.4 & \underline{3.5} & 2.6 & 1.5\\
2 & 100K & 5B & \underline{12.0M} & 3.7M & 646.8K (17\%) & 119.7K (19\%)   & 0.2 & \underline{0.3} & 0.2 & <0.1\\
 & 1M & 500B & 119.8M & \underline{137.0M} & 55.9M (41\%) & 11.9M (21\%)    & 2.0 & \underline{2.6} & 2.0 & <0.1\\
 & 10M & 50000B & 1.2B & \underline{11.3B} & 5.5B (49\%) & 1.2B (22\%)        & 20.0 & 33.5 & \underline{165.9} & 84.0\\
3 & 100K &  5B & \underline{57.7M} & 53.9M & 9.1M (17\%) & 1.9M (21\%)      & 1.1 & \underline{1.3} & 0.9 & 0.3\\
 & 1M & 500B & 577.1M & \underline{3.7B} & 898.4M (24\%) & 193.4M (22\%)    & 10.5 & 14.7 & \underline{35.2} & 19.3\\
\hline
\end{tabular}
\caption{Scaling behavior of symmetric deletion lookup with input dataset size and threshold distance using data from \cite{Emerson2017}. The table indicates the number of entries considered at different stages of XTNeighbor, as well as the time spent in the different streams of the algorithm. $\texttt{AllPairs}$ indicates the total number of pairs that would need to be considered by a brute-force approach, which is provided as a reference. \texttt{DeletionVariants} denotes the total number of deletion variants, \texttt{MatchedPairs} the number of matched deletion-variant pairs, \texttt{UMatchedPairs} the number of matched pairs retained after deduplication, while \texttt{Outputs} is the number of returned neighbors after threshold validation. The underlined numbers denote the maximum number in each row and category. The numbers in brackets represent the percentage of matched pairs retained in deduplication and threshold validation filtering steps. S1-4 refer to Stream 1-4 mentioned in Fig. \ref{fig:xt_neighbor_streaming}.}
\label{tab:in-depth}
\end{table}

The table highlights the large reduction in the number of sequence pairs the algorithm needs to consider compared to a brute-force approach across all parameter settings. The table also shows how for large input sizes or threshold choices the number of matched pairs exceeds the number of deletion variants as the former grows superlinearly. This becomes the key bottleneck at scale which the algorithm needs to overcome. Correspondingly, Stream 3, which generates all possible matched pairs for candidate sequences, and Stream 4, the final threshold validation step, take up a larger fraction of total runtime. Interestingly, the fraction of unique matched pairs retained after threshold validation does not decrease with input size or threshold, remaining stable at $\sim$20\% for $d \geq 2$ and growing with input size at $d=1$. The increased runtime of these steps at larger $d$ is thus not driven by a growing proportion of false positives, but instead scales in proportion to the number of true returned pairs. The fraction retained after deduplication, by contrast, does decrease with threshold (from 67\% at $d=1$ to 24\% at $d=3$ for $10^6$ sequences), reflecting that at larger thresholds a given pair is typically matched through several shared deletion variants.

\section{A sequence representation that lower-bounds edit distance}
\label{seqtovec}

Converting sequences to vectors enables the use of highly-optimized algorithms for vector-based similarity search. To ensure comparable output, we introduce a "bag-of-amino-acids" representation, for which the Euclidean distance in representation space puts a lower bound on edit distance (see proof below). Combined with the k-d tree algorithm \cite{kdtree_orig} for neighbor-finding this representation provides an example of how embeddings can be used, while retaining the ability to perform exact Levenshtein distance radius search.

\begin{figure*}[h]
    \centering
    \includegraphics[width=\textwidth]{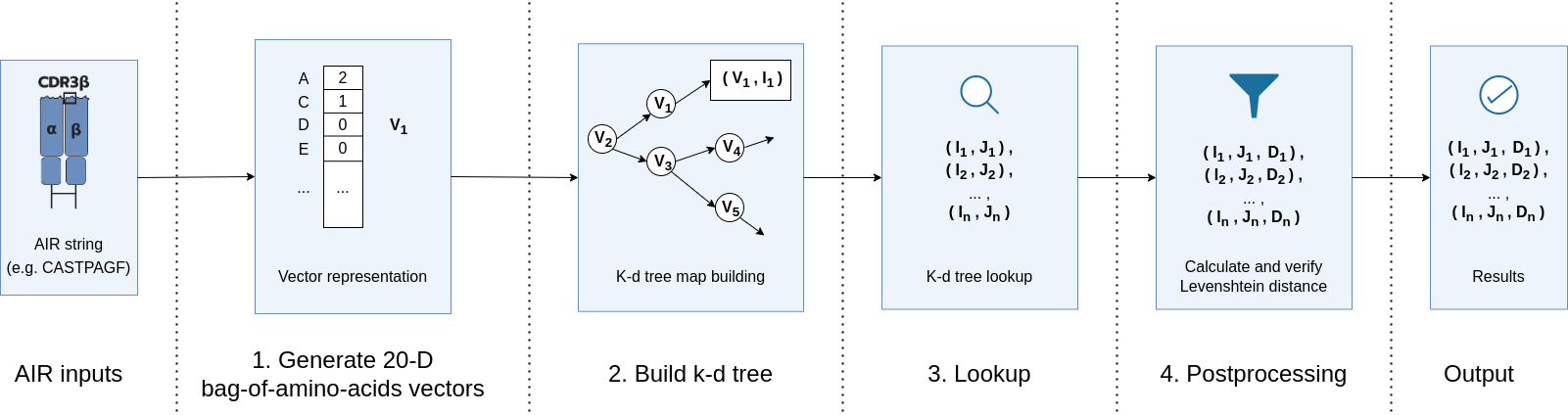}
    \caption{Process diagram of the k-d tree bag-of-amino-acids algorithm.}
    \label{fig:msc_diagram}
\end{figure*}

The representation is a 20-dimensional vector, where each element records the number of amino acids of each type in the sequence. As demonstrated below, whenever two sequences have an edit distance of $d$, their vector counterparts have an Euclidean distance $\leq \sqrt{2}d$ in this embedding space. Using this property, we can utilize k-d trees to identify nearest neighbors within a $\sqrt{2}d$ radius in representation space. The bag-of-amino-acids representation discards all positional information, which will result in false positives, that is, sequences with small embedding distance but larger edit distance. These sequence pairs need to be filtered out in a second processing step. An overall schematic of this approach is provided in Fig.~\ref{fig:msc_diagram}.

\bigskip

\begin{theorem}[Lower-bound inequality]
Let $x$ be the bag-of-amino-acids vector representation of a sequence, and $y_d$ the bag-of-amino-acids vector representation of a sequence that differs from $x$ by $d$ edits. \\
Then: $||x-y_d|| \leq \sqrt{2}d$
\end{theorem}

This theorem can be proved by induction. Let $i$, $j$ be any two orthogonal unit vectors in 20 dimensions representing an addition or deletion of an amino acid.

\raggedright Initial Case (d = 1):\\
{\centering
\begin{equation*}
y_1 = \begin{cases}
x+i &\text{for insertion}\\
x-i &\text{for deletion}\\
x+i-j &\text{for substitution}\\
\end{cases}
\end{equation*}
\begin{equation*}
\lVert x-y_1 \rVert = \begin{cases*}
\lVert x-(x \pm i) \rVert = \lVert \pm i \rVert = 1 \leq \sqrt{2} & \text{for insertion and deletion}\\
\lVert x-(x+i-j) \rVert = \lVert j-i \rVert = \sqrt{2} & \text{for substitution}\\
\end{cases*}
\end{equation*}
\par}

\raggedright Induction Step:\\
{\centering
\begin{equation*}
y_{d+1} = \begin{cases}
y_d+i &\text{for insertion}\\
y_d-i &\text{for deletion}\\
y_d+i-j &\text{for substitution}\\
\end{cases}
\end{equation*}

\begin{equation*}
\lVert x-y_{d+1} \rVert = \begin{cases}
\lVert x-(y_d \pm i) \rVert & \text{for insertion and deletion}\\
\lVert x-(y_d+i-j) \rVert & \text{for substitution}\\
\end{cases}
\end{equation*}
\par}
\begin{equation*}
\end{equation*}
\raggedright From the triangle inequality it follows:\\
{\centering
\begin{equation*}
\lVert x-y_{d+1} \rVert \leq \begin{cases}
\lVert x-y_d \rVert + \lVert \pm i \rVert \leq \sqrt{2}d + 1 < \sqrt{2}(d+1) & \text{for insertion and deletion}\\
\lVert x-y_d \rVert + \lVert j-i \rVert \leq \sqrt{2}d + \sqrt{2} = \sqrt{2}(d+1) & \text{for substitution}\\
\end{cases}
\end{equation*}
\par}

\clearpage

\section{Step-by-step procedures of each algorithm}
\label{app_sup_methods}

\begin{figure*}[b]
    \centering
    \includegraphics[width=\textwidth]{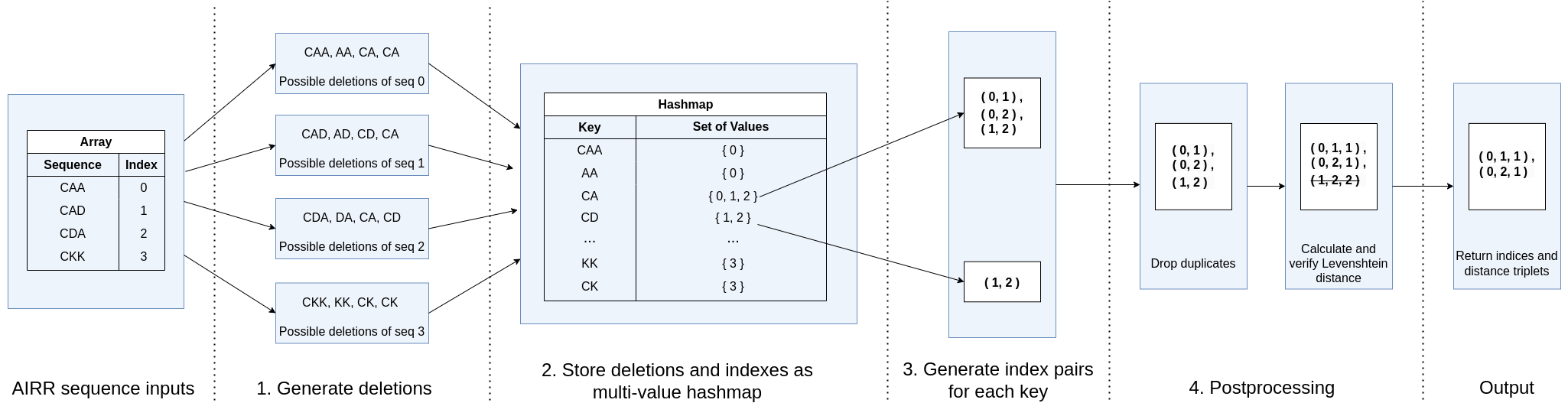}
    \caption{Process diagram of SymDel lookup algorithm with 4 example inputs and distance threshold 1.}
    \label{fig:symspell}
\end{figure*}

\subsection{SymDel lookup (Fig. \ref{fig:symspell})}
\begin{enumerate}
    \itemsep0em
    \setcounter{enumi}{-1}
    \item Given an array $\{s_i\}$ of AIR sequences and a distance threshold.
    \item Generate all deletion variants of the sequences $\{(v_{ij},i)\}$ up to the distance threshold.
    \item Store $\{v_{ij}\} \to \{i\}$ in a multi-value hash map.
    \item For each $(v_{ij},i)$, look up the matched sequence index $i'$ in the hash map using $v_{ij}$ as the search key.
    \item Deduplicate $\{(i,i')\}$.
    \item Calculate the sequences' distance, filter for threshold validity, then return $\{(i,i',d(s_i,s_{i'}))\}$.
\end{enumerate}

\subsection{SymScan}
\begin{enumerate}
    \itemsep0em
    \setcounter{enumi}{-1}
    \item Given an array $\{s_i\}$ of AIR sequences and a distance threshold.
    \item Generate an array $\{(v_{ij}, s_i)\}$ of all possible two-tuples describing a pair of an input sequence $s_i$ and a possible deletion variant $v_{ij}$ up to the distance threshold.\footnote{In the actual implementation, the two-tuples are actually $(h_{ij}, i)$ such that the first element is the hash $h_{ij}$ of the deletion variants instead of the variant strings $v_{ij}$ themselves, and the second element is the index $i$ of the input sequence instead of the actual sequence $s_i$.}\footnote{This step is parallelized.\label{fn:parallel}}
    \item Sort the two-tuple array\textsuperscript{\ref{fn:parallel}} according to the deletion variants $v_{ij}$, and deduplicate.
    \item Use the sorted array to identify groups of input strings $\{\{s_i\}\}$ that generate identical deletion variants, and transform this to an array of neighbor candidates $\{(s_i, s_{i'})\}$.\textsuperscript{\ref{fn:parallel}}
    \item For each candidate $(s_i, s_{i'})$, measure the distance between them, filter out false positives, to get an array of three-tuples $\{(s_i, s_{i'}, d_i)\}$ where $d_i$ is the distance between the two strings.\textsuperscript{\ref{fn:parallel}}
    \item Return the result.
\end{enumerate}

\begin{figure*}[p]
    \centering
    \includegraphics[width=\textwidth]{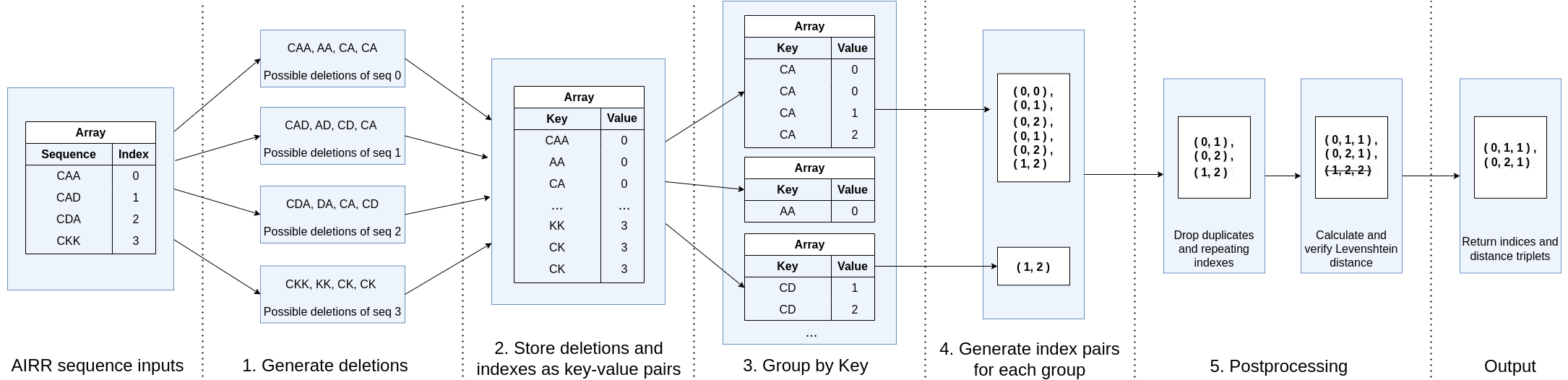}
    \caption{Process diagram of the XTNeighbor base algorithm with 4 example inputs and distance threshold 1.}
    \label{fig:xt_neighbor_base}
\end{figure*}

\subsection{XTNeighbor base algorithm (Fig. \ref{fig:xt_neighbor_base})}

The base algorithm is described here to explain the parallel design in isolation. It is superseded by the streaming implementation described below, which is the version used for all results in this work.

\begin{enumerate}
    \itemsep0em
    \setcounter{enumi}{-1}
    \item Given an array $\{s_i\}$ of AIR sequences and a distance threshold.
    \item Generate all deletion variants of the sequences $\{(v_{ij},i)\}$ up to the distance threshold.
    \item Group $\{(v_{ij},i)\}$ by $v_{ij}$ using merge sort and Run-Length Encoding algorithm \cite{run-length-encoding}.
    \item For each group $v_{g}$, generate all possible matching sequence pairs $\{(i',i'')\}$ within the group.
    \item Deduplicate $\{(i',i'')\}$ using merge sort and Run-Length Encoding algorithm.
    \item Calculate the sequences' distance, filter for threshold validity, then return $\{(i',i'',d(s_{i'},s_{i''}))\}$.
\end{enumerate}

\begin{figure*}[p]
    \centering
    \includegraphics[width=\textwidth]{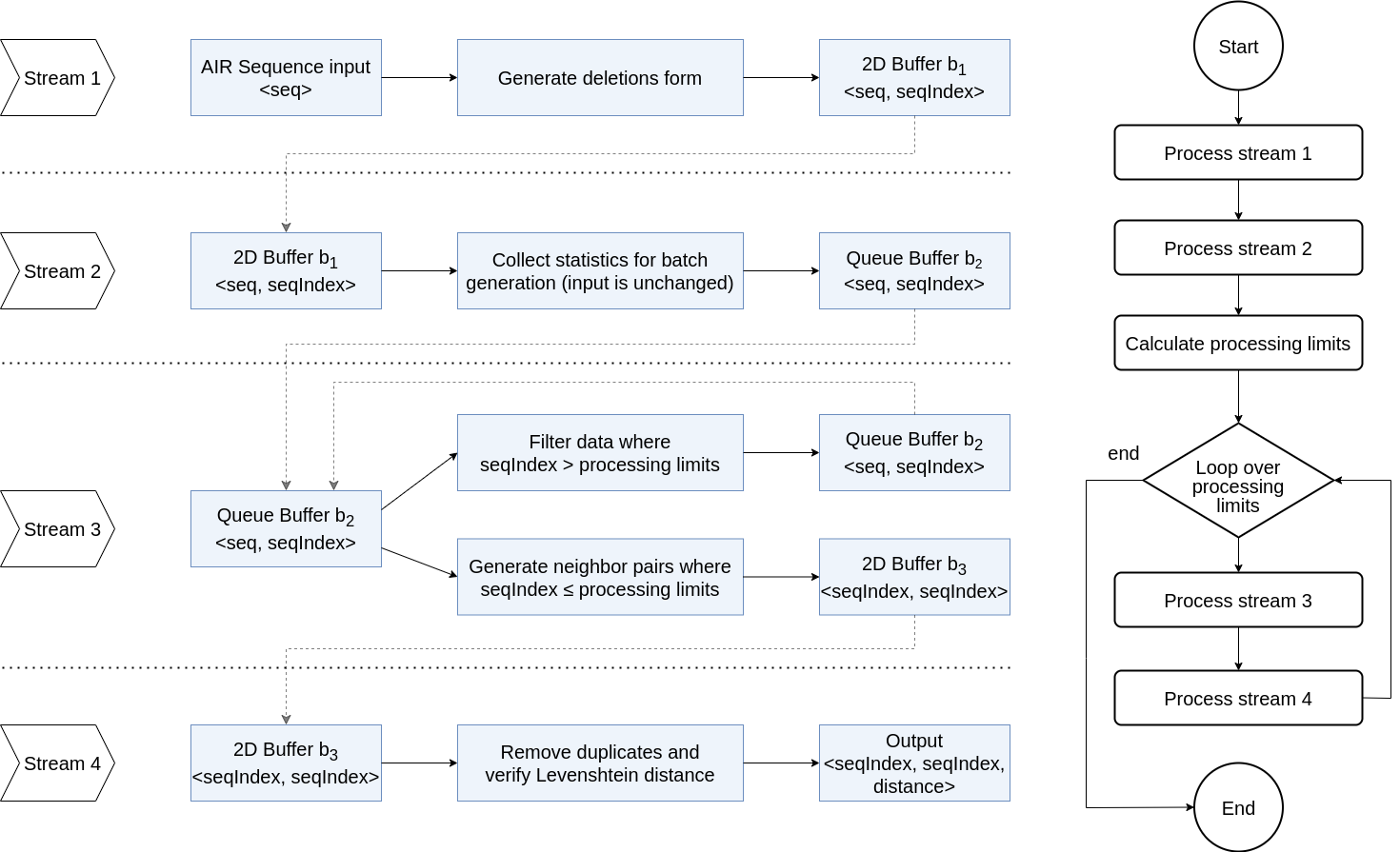}
    \caption{Process diagram of the XTNeighbor algorithm.}
    \label{fig:xt_neighbor_streaming}
\end{figure*}

\subsection{XTNeighbor (Fig. \ref{fig:xt_neighbor_streaming})}

\begin{enumerate}
    \itemsep0em
    \setcounter{enumi}{-1}
    \item Given an array $\{s_i\}$ of AIR sequences and a distance threshold.
    \item Initialize 2D buffers $b_1, b_3$ and queue buffer $b_2$. (See appendix \ref{app:streaming} for information on 2D buffer.)
    \item Stream from $\{s_i\}$ into $b_1$ with the following processors:
    \begin{enumerate}
         \item Generate all deletion variants of the sequences $\{(v_{ij},i)\}$ up to the distance threshold.
         \item Write $\{(v_{ij},i)\}$ to $b_1$.
    \end{enumerate}
    \item Stream from $b_1$ into $b_2$ with the following processors:
    \begin{enumerate}
        \item Group $\{(v_{ij},i)\}$ by $v_{ij}$ using merge sort and Run-Length Encoding algorithm \cite{run-length-encoding}.
        \item For each group $v_{g}$, generate all possible matching sequence pairs $\{(i',i'')\}$ within the group.
        \item Collect the histogram of $min(i',i'')$ for each group.
        \item Write the original unmodified input $\{(v_{ij},i)\}$ into $b_2$.
    \end{enumerate}
    \item Solve bin packing problem \cite{bin_packing} for processing limits $l_k$ from the histogram of $min(i',i'')$ using Next-fit algorithm \cite{next_fit}. This optimization maximizes the number of buffer columns that can be processed in parallel given the memory limit while ensuring that all duplicated results are processed in the same loop.
    \item Loop over the precalculated processing limit $l_k$.
    \begin{enumerate}
        \item Stream from $b_2$ conditionally into either $b_2$ or $b_3$.
        \begin{enumerate}
            \item Group $\{(v_{ij},i)\}$ by $v_{ij}$ using merge sort and Run-Length Encoding algorithm.
            \item For each group $v_{g}$, generate all possible matching sequence pairs $\{(i',i'')\}$ within the group.
            \item Filter $\{(i',i'')\}$ where $min(i',i'') \le l_k$, sort it by $min(i',i'')$, then save to $b_3$. This step ensures that all duplicated pairs are bounded within the same buffer column.
            \item Filter $\{(v_{ij},i)\}$ where $i > l_k$, then write back to $b_2$ for future processing.
        \end{enumerate}
        \item Stream from $b_3$ to output stream.
        \begin{enumerate}
            \item Deduplicate $\{(i',i'')\}$ using merge sort and Run-Length Encoding algorithm.
            \item Calculate the sequences' distance, filter for threshold validity, then return $\{(i',i'',d(s_{i'},s_{i''}))\}$.
        \end{enumerate}
    \end{enumerate}
\end{enumerate}

\begin{figure*}[p]
    \centering
    \includegraphics[width=0.35\textwidth]{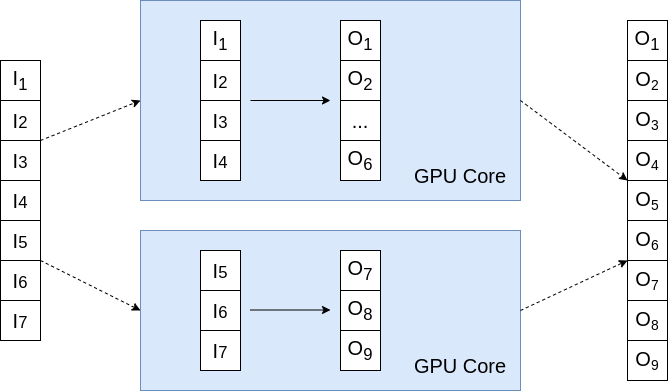}
    \hspace{0.5cm}
    \includegraphics[width=0.5\textwidth]{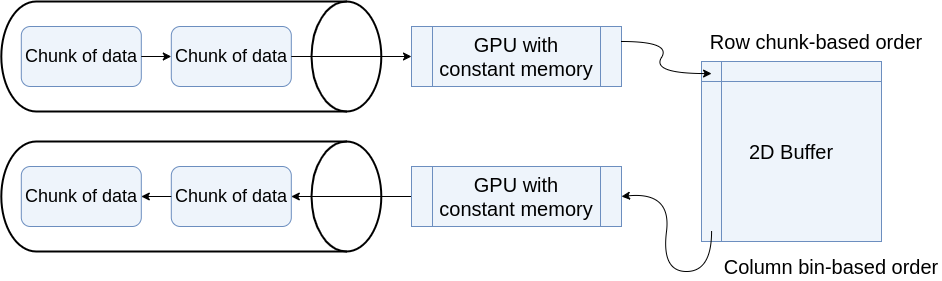}
    \caption{The data flow diagram related to (left) the parallel processing of group-by operations that avoid race condition by preallocating the outputs and (right) 2-dimensional buffer for changing the direction and grouping of data stream.}
    \label{fig:data_flow}
\end{figure*}

\subsection{Writing and reading the 2-dimensional buffer used for streaming in XTNeighbor (Fig. \ref{fig:data_flow})}

Writing:
\begin{enumerate}
    \itemsep0em
    \setcounter{enumi}{-1}
    \item Given a chunk of data $D$ which is stored as array, 2D buffer $B$, number of bins $N$, and current available GPU memory $M$.
    \item Split each element in $D$ into $N$ equal bins as $D_1,D_2,...,D_N$.
    \item Append a row to buffer $B$ with $B_{ki} = D_i$ for $i=1,2,...N$ and $k$ denotes the index of the last row.
    \item Update the memory size $S_i$ of each column for further use.
    \item Once the last row is written, solve Next-fit Bin optimization to determine the number of column $n_j$ to read for the $j^{th}$ read operation using $M, S_i$.
\end{enumerate}

Reading: 
\begin{enumerate}
    \itemsep0em
    \setcounter{enumi}{-1}
    \item Given a 2D buffer $B$ and number of columns to read $n_j$ which is calculated on the last writing operation.
    \item Read and return $B_{ki}$ for $i=1,...,n_j$ and every row $k$.
    \item Remove the used columns.
\end{enumerate}

\clearpage

\clearpage

\end{document}